\documentclass[apl,twocolumn, amsmath,amssymb]{revtex4}

\usepackage{graphicx}
\usepackage{dcolumn}
\usepackage{bm}

\begin{document}

\title{Selective epitaxial growth of sub-micron complex oxide structures by amorphous SrTiO$_3$}

\author{P. Morales}
 \email{patrick.morales@utoronto.ca}
\author{M. DiCiano}%
\author{J.Y.T. Wei}
 \affiliation{Department of Physics, University of Toronto, 60 St. George Street, Toronto, ON M5S1A7, Canada}


\begin{abstract}
A chemical-free technique for fabricating submicron complex oxide
structures has been developed based on selective epitaxial growth.
The crystallinity and hence the conductivity of the complex oxide is
inhibited by amorphous SrTiO$_3$ (STO). Using a combination of
pulsed laser deposition and electron-beam lithography, amorphous STO
barriers are first deposited on a single crystal substrate. A thin
film is then deposited on the patterned substrate with the amorphous
STO barriers acting to electrically and physically isolate different
regions of the film.  Since no chemical or physical etchants come in
contact with the deposited film, its integrity and stability are
preserved. This technique has successfully produced sub-micron
YBa$_2$Cu$_3$O$_{7-\delta}$ and La$_{2/3}$Ca$_{1/3}$MnO$_3$
structures.
\end{abstract}

\maketitle

Fabrication of complex oxide nanostructures is important from both a
physical and technological standpoint.  Many complex oxides exhibit
interesting electrical behavior including high-\emph{T$_c$}
superconductivity, colossal magnetoresistance (CMR) and
ferroelectricity. Potential device applications including high
sensitivity sensors,\cite{Braginski-FED1:1990} flux flow
transistors,\cite{Martens-IEEE1:1991} ferroelectric field effect
transistors\cite{Ramesh-SCI252:1991} and magnetic
memory\cite{Grishen-APL74:1999} require suitable microfabrication
techniques. However, due to inherent stoichiometric and structural
complexities associated with complex oxides, producing high quality
sub-micron structures has proven difficult because conventional
methods tend to require etching of the complex oxide.

\begin{figure}[h]
\centering
\includegraphics[width=8.5cm]{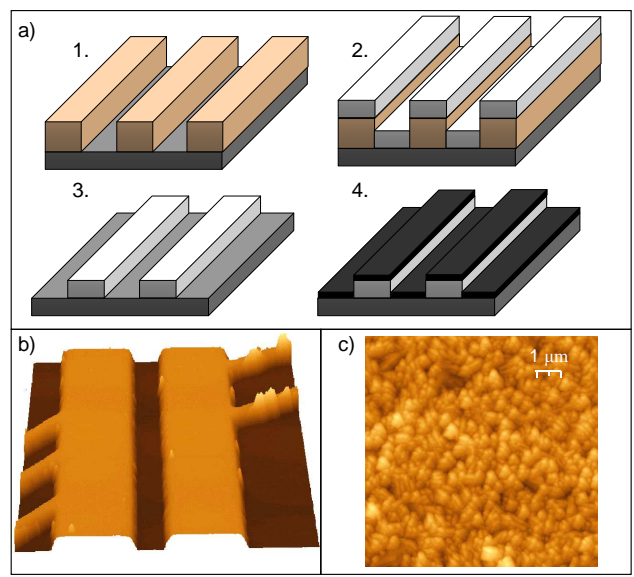}
\caption{\label{Technique} a$)$ Selective Epitaxial Nanofabrication
Technique: 1. A PMMA mask is defined using E-beam lithography. 2. A
layer of amorphous STO is deposited by PLD at ambient temperature.
3. The PMMA mask is removed leaving only the amorphous STO deposited
directly on the substrate. 4. A complex oxide thin film is deposited
on the patterned substrate. The material deposited on top of the
amorphous STO barriers is of poor crystallinity and hence poor
conductivity. b$)$ 3D AFM image of a typical patterned substrate
showing the amorphous STO barriers. c$)$ AFM image of a 50 nm YBCO
thin film deposited on the amorphous STO barrier, showing the
amorphous nature of the deposited YBCO with an average $($rms$)$
surface roughness of 19 $\pm$ 2 nm.}
\end{figure}

In this letter we present a technique for fabricating sub-micron
complex oxide structures which requires no chemical or physical
etching. Based on selective epitaxial growth, a single crystal
substrate is patterned both vertically and laterally by the
deposition of amorphous SrTiO$_3$ (STO) barriers. The amorphous
nature of the STO barriers acts to ensure that any material
deposited on top of the barriers is electrically insulating.  This
technique has the advantage that complex structures can be
fabricated without any degradation due to chemical or physical
etching. The minimum dimensions of the microstructure are well
defined by the amorphous STO barriers and are not determined by
diffusive processes. This technique also allows for the deposition
of passivation layers which can improve the stability of the complex
oxide.\cite{Copetti-APL61:1992}

\bigskip

Conventional techniques for fabricating sub-micron complex oxide
structures involve either post-deposition etching of the oxide or
patterning of the substrate before deposition. Post-deposition
patterning has commonly been obtained using wet-chemical etching of
the oxide in solutions of bromine in
ethanol,\cite{Vasquez-APL53:1988} ethylenediaminetetraacetic acid
(EDTA),\cite{EDTA} phosphoric acid\cite{Lyons-IEEE27:1991} or
hydrofluoric acid.\cite{Eidelloth-APL59:1991}  However, wet chemical
etching of YBa$_2$Cu$_3$O$_{7-\delta}$ (YBCO) thin films has shown
to cause a significant increase in the high-frequency surface
resistance as well as to cause insulating dead layers and a change
in the surface morphology at the exposed surfaces.
\cite{Roshko-IEEE5:1995} Physical etching methods such as reactive
ion etching \cite{RIE}, focused ion beam etching \cite{FIB}, and
pulsed laser etching \cite{Inam-APL51:1987} have also been used to
pattern complex oxide microstructures. However, these methods can
cause physical damage to the exposed surfaces of the thin film.
Also, heat generated from these physical processes can be
sufficiently large to alter the doping and hence their electrical
transport properties of the oxide.

Sub-micron structures can also be fabricated by patterning a single
crystal substrate prior to deposition.  The substrate is patterned
such that regions of the deposited film are physically separated
through the creation of ridge or trench structures
\cite{Mohanty-PHYC408:2004} or electrically isolated through the
inhibition of the conductivity of select regions. The inhibition of
conductivity can be achieved by destroying the local crystallinity
of select regions.  This can be achieved by diffusion of Si
\cite{Ma-APL55:1989}, SiO$_2$ \cite{Copetti-APL61:1992}, Si$_x$N$_y$
\cite{Kern-JVSB9:1991} or Ti \cite{Rossel-PHYC185:1991} or by
selectively determining where epitaxial growth can occur through the
deposition of a Ti or W layer on selective regions of the substrate.
\cite{Damen-SST11:1998}

\bigskip

Our chemical-free procedure for fabricating complex sub-micron
structures is outlined schematically in figure 1.  First, a
polymethylmethacrylate (PMMA) mask is defined on a single crystal
STO substrate by electron-beam lithography (EBL).  A $\approx$700 nm
layer of 7\% PMMA in anisole with a molecular weight of 450 amu is
spun onto a single crystal STO substrate at 4500 rpm for 45 s.  The
resist is then baked on a hotplate at 180$^o$C for 5 minutes. The
resist is then exposed by EBL using a JEOL IC-848A, tungsten
filament scanning electron microscope at an accelerating voltage of
30 kV, probe current of 8 pA and a line dosage of 0.9 nC/cm.  The
low probe current was used to allow sufficient time for any
accumulated charge to dissipate, in order to reduce any electron
charging effects that could readily occur when imaging an insulating
material.  The resist is then developed in a 3:1 solution of
isopropyl alcohol (IPA) to methyl isobutyl ketone (MIBK) for 30
seconds and then rinsed in IPA.

After the mask has been defined, a layer of amorphous STO is then
deposited by pulsed laser deposition (PLD) using a 248 nm KrF
excimer laser at a laser energy density of $\approx$1.2 J/cm$^2$.
The STO is pulsed laser ablated at an oxygen partial pressure of 20
mTorr and at ambient temperature to ensure no evaporation or
diffusion of the PMMA mask occurs. An ablation time of 30 min using
a pulsed laser repetition rate of 10 Hz results in the deposition of
$\approx$350 nm of amorphous STO. The remaining PMMA is then removed
by acetone in an ultrasonic bath for 45 s, leaving behind an
inhibitive pattern of amorphous STO.  A complex oxide film, such as
YBa$_2$Cu$_3$O$_{7-\delta}$ (YBCO) or La$_{2/3}$Ca$_{1/3}$MnO$_3$
(LCMO) can then be deposited by PLD using typical ablation
conditions. Since the deposition of the complex oxide is the last
step in the process, the integrity of the resulting structure is
preserved as it does not undergo subsequent heating due to physical
bombardment or come in contact with chemical etchants.

\begin{figure}[h]
\centering
\includegraphics[width=8.5cm]{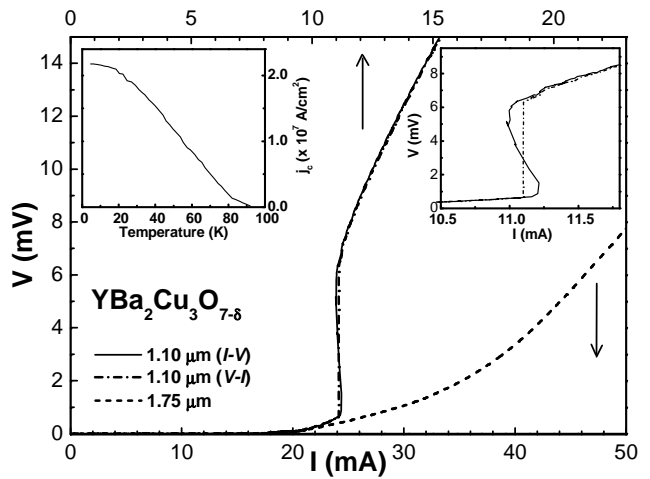}
\caption{\label{YBCO_IV}Current-Voltage characteristics of YBCO
microstrips of differing widths at 24 K.  The $V$-$I$ of the
narrower (1.10 $\mu$m) microstrip exhibits a characteristic step.
The $I$-$V$ measurement of the strip shows that this step coincides
with an $s$-shaped conductance region (inset upper right) both of
which are consistent with phase-slip behavior.  A $V-I$ measurement
of a wider YBCO strip does not exhibit such characteristic behavior.
Upper left inset shows the critical current density versus
temperature of the 1.10$\mu$m YBCO strip.}
\end{figure}

The electrical transport properties of our fabricated
microstructures were characterized by standard four-point, phase
sensitive, AC resistance versus temperature measurements, as well as
synchronous pulsed current versus voltage ($I$-$V$) and voltage
versus current ($V$-$I$) measurements. In the latter technique,
pulses of 200 $\mu$s with a duty cycle of 5\% were used in order to
minimize any effects of Joule heating.

Sub-micron YBCO strips fabricated using our technique exhibit
superconducting critical temperatures, $T_c$, from 80-92 K, with
transition widths of 1-10 K. YBCO strips show room temperature
resistivities of tens to several hundreds of $\mu\Omega\cdot$cm and
critical current densities, j$_c$, on the order of 10$^7$ A/cm$^2$,
which approach those of unpatterned YBCO thin films. To ensure that
an applied current was indeed confined to the selective regions of
the film, we also measured the YBCO film that was deposited on the
amorphous STO barriers. These control samples show no
superconducting transition and exhibit room temperature
resistivities on the order of 10$^{6}$ $\mu\Omega\cdot$cm. An atomic
force micrograph (AFM) shown in figure 1 confirms the amorphous
nature of the deposited YBCO.

\begin{figure}[h]
\centering
\includegraphics[width=8.5cm]{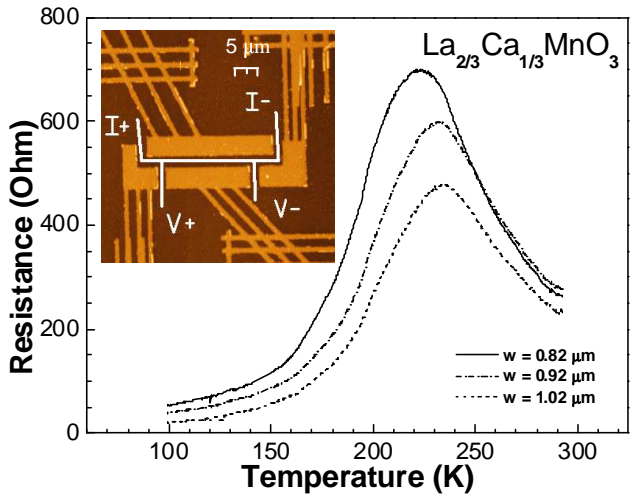}
\caption{\label{LCMO} Resistance versus temperature measurements for
LCMO microstrips of differing widths.  Inset shows an AFM image of a
typical microstrip with an overlay indicating the current and
voltage leads.}
\end{figure}

Previous transport studies on chemically-patterned cuprate samples
ranging from 100 nm to 2 $\mu$m in width have been reported. Highly
non-linear $I$-$V$ characteristics and anomalous resistance versus
temperature behavior have been attributed to either collective
flux-flow, phase slips or mesoscopic domains
\cite{Rogalla,Dmitriev,Jelila,AbdelhadiJung,Bonetti}.  The $V$-$I$
characteristics of YBCO microstrips fabricated using our technique
are shown in Figure 2.  For wider strips, the $V$-$I$ characteristic
can be described by a power law relationship caused by thermally
activated flux creep and flux flow.  However, as the applied current
becomes more confined in a narrower strip, the power law relation is
only valid below some threshold value of applied current. At this
threshold value, a discontinuity in the voltage occurs and the $V-I$
relationship becomes linear.  An $I$-$V$ measurement of the same
microstrip shows that voltage discontinuity coincides with an
$s$-shaped, negative conductance region (see inset Figure 2). Both
the discontinuity of the voltage in the $V$-$I$ characteristic and
the $s$-shaped $I$-$V$ characteristic of the microstrip are
consistent with phase-slip behavior due to the lateral confinement
of the superconductor. \cite{SBT, Vodolazov}  An investigation of
the phase-slip behavior in our YBCO microstrips as well as
information derived from such behavior on the dynamics of the
superconducting order parameter is the focus of separate papers.
\cite{Morales-NanoYBCO, Morales-Tau_Psi_Phi}

Sub-micron LCMO strips were also fabricated using our technique.
Sub-micron LCMO strips are of interest because their
magnetoresistive properties can be tuned by externally applied
strain. \cite{Koo-APL71:1997} Resistance versus temperature
measurements of strips of differing widths at zero field are shown
in figure 3.  Fabricated sub-micron LCMO strips show a maximum in
resistance, $T_m$, ranging from 220 K to 300 K.  The position of
$T_m$ has been shown to be highly dependant on stress relaxation and
improved crystallinity resulting from grain growth.
\cite{Thomas-JAP84:1998} A field dependent study of the CMR
properties of sub-micron LCMO strips fabricated using our technique
will be presented in a separate paper. \cite{Morales-LCMO}

\bigskip

In summary, we have developed a novel chemical-free technique for
fabricating sub-micron complex oxide structures. The technique is
based on selective epitaxial growth of a complex oxide thin film.
The crystallinity and hence the conductivity of the complex oxide
thin film are inhibited by amorphous SrTiO$_3$ barriers deposited
upon a single crystal substrate. This technique has been
successfully applied to fabricate strips of sub-micron
high-\emph{T$_c$} superconducting YBa$_2$Cu$_3$O$_{7-\delta}$ and
colossal magnetoresistive La$_{2/3}$Ca$_{1/3}$MnO$_3$.

\bigskip

The authors acknowledge assistance by Stephanie Chiu and Eugenia Tam
and funding from: NSERC, CFI/OIT, MMO/EMK and the Canadian Institute
for Advanced Research under the Quantum Materials Program.

\end{document}